\documentclass[aps,prl,showpacs,twocolumn]{revtex4-1}
\usepackage{amssymb}
\usepackage{graphicx}
\usepackage{amsmath}
\usepackage{color}

\def\bk{{\bf k}}

\def\br{{\bf r}}
\def\bR{{\bf R}}
\input{epsf}
\begin{document}

\title{Strong quantum interference in strongly disordered bosonic insulators}
\author{S.V.~Syzranov$^{1,2}$, A.~Moor$^1$, K.B.~Efetov$^1$}
\affiliation{$^1$Theoretische Physik III, Ruhr-Universit\"at Bochum, 44780 Bochum, Germany, \\
$^2$Institute for Theoretical Condensed Matter Physics,
   Karlsruhe Institute of Technology,
   76131 Karlsruhe, Germany}
\date{\today}

\begin{abstract}
We study the variable-range hopping (VRH) of bosons in an array of
sites with short-range interactions and a large characteristic
coordination number. The latter leads to strong quantum interference
phenomena yet allows for their analytical study. We develop a
functional renormalization group scheme that repeatedly eliminates
high-energy sites properly renormalizing the tunnelling between the
low-energy ones. Using this approach we determine the temperature
and magnetic field dependence of the hopping conductivity and find a
large positive magnetoresistance. With increasing magnetic field the
behaviour of the conductivity crossovers from the Mott's law to an
activational behaviour with the activation gap proportional to the
magnetic field.
\end{abstract}

 \maketitle












Transport in a huge variety of insulators is described by the model
of variable-range hopping (VRH)\cite{Mott:book} based on a few
rather general
assumptions: charge carriers hop inelastically 
between random-energy sites, e.g. impurities, grains, or (quasi-)localized
states, the probability of a hop exponentially decreases with distance and
has an activational temperature dependence. The particular nature of the
sites and the bath is not important,-- the model can be applied to the
transport of electrons in doped semiconductors as well as to that of Cooper
pairs in macroscopically insulating superconductive materials.

In the course of each inelastic hop between two sites a charge
carrier may tunnel elastically {via} intermediate sites. The
interference between different-{path} contributions gives rise to
mesoscopic and magneto-transport phenomena\cite{Shklovskii:review}.
The interference is suppressed in strongly insulating materials
{because there} the elastic transmission is dominated either by the
direct tunnelling path between two sites or by the shortest chain of
intermediary sites. With increasing the intersite couplings or the
site concentration, more alternative paths come into play and the
interference effects get more pronounced.

This Letter is devoted to the study of the VRH transport in bosonic
insulators with large coordination numbers. In these materials the
interference phenomena are particularly strong because of the
constructive interference between nearly all tunnelling processes,
including the many-body ones. Large coordination number $Z\gg1$ of
the effective array of sites allows us to treat the interference
effects analytically, unlike the previous VRH studies, and is
characteristic of a broad class of insulating materials. Those
include, for example, superconductive films close to the
metal-insulator transition (see, e.g., Ref.~\cite{Feigelman:rubbish}
for a review), densely packed granular materials, disordered bosonic
systems with many particles on the localization length,
``superconductive hay''\cite{Vinokur:hay}, etc. Strong interference
in these materials manifests itself, for instance, in a huge
magnetoresistance\cite{GMR},-- magnetic field may change conductance
by orders of magnitude or even drive a superconductor into a strong
Cooper-pair insulator.
 Transport phenomena and
superconductor-insulator transitions are also being observed since
recently in artificially made arrays of Josephson junctions with
$Z\gg1$\cite{Gershenson}.

In general, the interference in these materials is intractable
perturbatively, unlike many earlier VRH models (cf., e.g.,
Ref.~\cite{Perturbative}). Tunnelling along paths of sites with close
energies\cite{Longinterference} is also irrelevant for such systems
due to the small portion of such paths among all possible ones.

\textit{Model.} Although we focus on the transport of bosons, the
results of this Letter apply also to systems with an arbitrary
charge carrier statistics if the VRH is dominated by single-particle
hops. The latter may be the case, for instance, in an electronic
system with a shallow Fermi sea, that is sufficiently exceeded by
the characteristic Coulomb energy. As we consider systems with
$Z\gg1$, we also expect the results to hold qualitatively in the
materials with coordination numbers of order unity.

We assume for simplicity {that} the intersite interactions
{are} short-range and the number of particles on each site 
uniquely determines their state. This simplification is possible if
the other states are absent or separated by large energy gaps, which
is characteristic of electrons in doped
semiconductors\cite{Shklovskii:book}, Cooper-pairs in
macroscopically insulating media\cite{Efetov:fundamental,
CayleySIT}, and of the charge carriers in nearly all other materials
under consideration.

The particle motion is governed by the Hamiltonian
\begin{equation}
\hat{\mathcal{H}}=\sum_{\mathbf{r}}U_{\mathbf{r}}(n_{\mathbf{r}})-\sum_{%
\mathbf{r}\neq \mathbf{r}^{\prime }}t_{\mathbf{r}\mathbf{r}^{\prime }}{\hat{b%
}}_{\mathbf{r}}^{\dagger }{\hat{b}}_{\mathbf{r}^{\prime }}.  \label{Haminit}
\end{equation}%
Here $U_{\mathbf{r}}(n_{\mathbf{r}})$ is the energy of $n_{\mathbf{r}}$
particles on site $\mathbf{r}$, a large random quantity affected by disorder
and stray charges close to the site. The last term in Eq.~(\ref{Haminit})
accounts for the intersite hopping, ${\hat{b}}_{\mathbf{r}}^{\dagger }$ and $%
{\hat{b}}_{\mathbf{r}}$ being (bosonic) creation and annihilation operators
on grain $\mathbf{r}$, $n_{\mathbf{r}}={\hat{b}}_{\mathbf{r}}^{\dagger }{%
\hat{b}}_{\mathbf{r}}$.

Below we count the energies of all excitations
from the energy of the ``ground charging state'', the configuration of
integers $n_{\mathbf{r}}$ that minimize the first term in the right-hand
side (rhs) of Eq.~(\ref{Haminit}).
Typically, the energies $E_{\mathbf{r}}^{+}$ and $%
E_{\mathbf{r}}^{-}$, needed respectively to add or remove a particle
from the site $\br$, are determined by the Coulomb interactions with
the other particles and stray charges, and we refer to these
energies and on-site excitation states as respectively ``charging''
energies and states.

As only low-energy states are relevant for low-temperature
conductance, one can conveniently rewrite the Hamiltonian in {a}
reduced Hilbert space,-- each site $\mathbf{r}$ has only 3 states:
$|\pm 1\rangle _{\mathbf{r}}$ and $|0\rangle _{\mathbf{r}}$,
corresponding respectively to $\pm 1$ and $0$ extra particles as
compared to the ground charging state. Introducing spin-$1$
operators; $S_{\mathbf{r}}^{z}|\pm 1\rangle _{\mathbf{r}}=\pm |\pm
1\rangle _{\mathbf{r}}$, $S_{\mathbf{r}}^{z}|0\rangle =0$,
$S_{\mathbf{r}}^{\pm }|\mp 1\rangle =\sqrt{2}|0\rangle $,
$S_{\mathbf{r}}^{\pm }|0\rangle =\sqrt{2}|\pm 1\rangle $, we arrive
at the reduced Hamiltonian
\begin{eqnarray}
\hat{\mathcal{H}} &=&\frac{1}{2}\sum_{\mathbf{r}}\left[ \left( E_{\mathbf{r}%
}^{+}+E_{\mathbf{r}}^{-}\right) \hat{S}_{\mathbf{r}}^{z}\hat{S}_{\mathbf{r}%
}^{z}+(E_{\mathbf{r}}^{+}-E_{\mathbf{r}}^{-})\hat{S}_{\mathbf{r}}^{z}\right]
\notag \\
&&-\frac{1}{2}\sum_{\mathbf{r},\mathbf{r}^{\prime }}J_{\mathbf{r}\mathbf{r}%
^{\prime }}\hat{S}_{\mathbf{r}}^{+}\hat{S}_{\mathbf{r}^{\prime
}}^{-}, \label{HamS}
\end{eqnarray}
where the exact form of the coupling constants $J_{\br\br^\prime}$
depends on the microscopic details of the model. For instance, for a
bosonic system with large occupancies $n_\br\gg1$ in the charging
ground state, $J_{\br\br^\prime}=(n_\br
n_{\br^\prime})^{1/2}t_{\br\br^\prime}$.

In absence of inter-site interactions at strong disorder, the
density of states (DoS) of the (anti)particle states has no Coulomb
gap\cite{Shklovskii:book} and, thus, reaches some finite value $\nu$
at low energies. The assumption of a constant DoS in a certain
energy interval is crucial for our consideration.

\textit{Site-decimating renormalization group.} The transport of low-energy excitations
can be analysed by constructing a logarithmic renormalization group
(RG) technique that repeatedly eliminates the highest-energy
charging states properly renormalizing the intersite tunnelling
couplings between all the lower-energy states.

Let us select a few (anti-)particle charging states with the very
highest energies and split the Hamiltonian into the Hamiltonian
$\hat{\mathcal{H}}_{high}$ of the respective high-energy charging
states, the Hamiltonian $\hat{\mathcal{H}}_{low}$ of the rest of the
system, and the tunnelling $\hat V$ between the selected states and
all the others. Any eigenstate of the system is, strictly speaking,
a superposition of the charging states on all sites. The projection
$\Psi_{low}$ of the eigenfunction on the lower-energy charging
states
satisfies Schr{\"o}dinger equation $\hat{\mathcal{H}}_{low}^\prime%
\Psi_{low}=\varepsilon\Psi_{low}$ with an effective (exact)
Hamiltonian
\begin{eqnarray}
\hat{\mathcal{H}}_{low}^\prime=\hat{\mathcal{H}}_{low}-\hat V({\hat{\mathcal{%
H}}_{high}-\varepsilon})^{-1}\hat V,  \label{Hmodif}
\end{eqnarray}
$\varepsilon$ being the eigenenergy. The latter may be neglected in
Eq.~(\ref{Hmodif}) if the energies of the selected states
significantly exceed the energies of the other states.

Assume, the intersite tunnelling elements are small, so that one can
neglect (anti-)particle hopping between the highest-energy charging
states. Then we find that the modified Hamiltonian~(\ref{Hmodif})
has the same form as the initial Hamiltonian~(\ref{HamS}) {provided
the couplings are renormalized},
\begin{equation}   \label{Jmod}
J_{\br\br^\prime}\rightarrow J_{\br\br^\prime} +
\sum_{\bR_p}\frac{J_{\br\bR_p}J_{\bR_p\br^\prime}}{E_{\bR_p}^+}
+\sum_{\bR_h}\frac{J_{\br\bR_h}J_{\bR_h\br^\prime}}{E_{\bR_h}^-},
\end{equation}
where $\bR_p$ and $\bR_h$ label the eliminated particle and
antiparticle states respectively. The increment of the magnitude of
the coupling $|J_{\br\br^\prime}|$ is positive, as the energies
$E_{\bR_p}^+$ and $E_{\bR_h}^-$ in Eq.~(\ref{Jmod}) are positive by
definition.

Eq.~(\ref{Jmod}) can be understood as follows. In addition to the
direct tunnelling between sites $\br$ and $\br^\prime$ a particle may
cotunnel via virtual high-energy states in one of two ways: 1)via
{an} intermediate particle state on site $\mathbf{R}$,
or 2)first create a particle-antiparticle dipole on sites $\mathbf{r}^{\prime },\mathbf{%
R}$, then---on sites $\mathbf{R}$, $\mathbf{r}$. The respective
high-energy particle or antiparticle state on the site $\bR$ can be
omitted provided the coupling constants $J_{\br\br^\prime}$ are
renormalized. Since processes 1) and 2) interfere constructively in
a bosonic system, the renormalization increases the tunnelling.


The generic Eq.~(\ref{Jmod}) may be of little use for arbitrary
distributions of the site energies and locations. However, in the
case of a large coordination number $Z$ the averaging over the site
positions leads, as we show below, to a simple expression involving
the averaged function $J_{\br\br^\prime}$. Moreover, large
coordination number suppresses the fluctuations of strongly
renormalized couplings, since all the decimated states contribute
with the same sign to each coupling, while the fluctuations result
from multiple random-sign contributions.

 Repeatedly eliminating the
highest-energy charging (anti-)particle states and introducing
continuous coordinates in the coupling function we arrive at a
functional RG equation with a running high-energy cutoff $E$:
\begin{eqnarray}
    \partial_l J_{\br\br^\prime}=\lambda\int
    J_{\br\bR}J_{\bR\br^\prime}d\bR, \quad
    l=\log({E_0}/{E}),
    \label{RG}
\end{eqnarray}
where the quantity $\lambda=2\nu$ is independent of energy, as we assumed
previously.

On each RG step we considered the cotunnelling via only one
intermediate site, which is justified for sufficiently small
couplings $J$,
\begin{equation}
\lambda J\equiv \int \lambda J_{\mathbf{r}\mathbf{r}^{\prime
}}d\mathbf{r}^{\prime }\ll 1.  \label{smallJ}
\end{equation}%
Indeed, taking into account the tunnelling between eliminated
charging states in the $n$-th order results in $\sim \left[ (\lambda
J)\log (E^{\prime }/E)\right] ^{n}$ relative correction to the
tunnelling couplings on an RG step with energy rungs $E^{\prime }$
and $E$. Thus, Eq.~(\ref{RG}) is the one-loop functional RG equation
for the intersite tunnelling elements.

Eq. (\ref{RG}) should be solved implying the coupling
$J_{\mathbf{rr}^{\prime }}$ corresponds to the bare one at the
energies of order of the characteristic charging energy $E_{0}$ in
the non-renormalized Hamiltonian (\ref{HamS}). The RG flow must be
stopped if, as a result of the renormalization, the couplings become
too large, violating condition (\ref{smallJ}), or if the high-energy
cutoff $E$ reaches either some characteristic energy of relevant
conducting excitations or the mean level spacing $1/(\nu \xi^{d})$
on the characteristic radius $\xi $ of the
renormalized function $J_{\mathbf{r}\mathbf{r}^{\prime }}$ in a $d$%
-dimensional space.

The smallness of the fluctuations of the coupling on each RG step
requires $Z(E/E^\prime)[\ln(E^\prime/E)]^2\gg1$. So long as the
fluctuations are suppressed, the renormalized function
$J_{\br\br^\prime}$ fully describes the conduction. We emphasize
that this is characteristic of a bosonic system, considered in the
present Letter, and makes our analysis very different from that of
fermionic insulators\cite{Spivak:negativeMR, Shklovskii:review}. In
a fermionic system our RG scheme would yield a non-renormalized
coupling function with large fluctuations, which would determine
then the transport properties of the system. The latter does not apply,
however, to a system with a so shallow Fermi sea that the transport
is effectively single-particle,-- the renormalization then occurs
similarly to the bosonic case.

\textit{Mean-field critical point.} In absence of magnetic field
the intersite tunnelling is translationally invariant, $J_{\mathbf{r}%
\mathbf{r}^{\prime }}=J_{\mathbf{r}-\mathbf{r}^{\prime }}$. Then for the
Fourier-transorm $J_{\mathbf{k}}$ of the coupling function Eq.~(\ref{RG})
yields
\begin{equation}
J_{\mathbf{k}}(l)={J_{\mathbf{k}}(0)}[{1-\lambda
J_{\mathbf{k}}(0)l}]^{-1}. \label{Jk}
\end{equation}%
where $J_{\mathbf{k}}\left(0\right)$ is the bare value of the
tunnelling amplitude. In the limit of a strong insulator, $\lambda
J_{\mathbf{k}}l\ll 1$, the renormalization results in a small
correction to the tunnelling and can also be obtained perturbatively
by considering the tunnelling processes via one intermediate site.

The renormalization is more pronounced in less insulating materials.
In particular, the renormalized tunnelling elements diverge if the
coordination number $Z$ or the bare coupling are sufficiently large,
namely, if the quantity $J_{\mathbf{k}}(0)$ exceeds some critical
value $J_{c}$,
\begin{equation}
\lambda J_{c}l_{max}=1,  \label{MFC}
\end{equation}%
$l_{max}$ being the maximal value of the logarithm that stops the RG flow.

Eq.~(\ref{MFC}) matches the mean-field (MF) criterion of the phase
transition in a system with Hamiltonian (\ref{HamS}) and the uniform
order parameter $\Delta =\langle \hat{S}_{\mathbf{r}}^{+}\rangle $.
However, we emphasize that the above analysis is not sufficient to
prove the existence of the metal-insulator or
superconductor-insulator transition when condition (\ref{MFC}) is
fulfilled, because we have to stop the RG flow if the inequality
(\ref{smallJ}) no longer applies. Clearly, to study such transitions
one has to employ alternative methods, which is beyond the scope of
this paper. Now, we concentrate on the transport in the insulating
regime when the effective amplitude $J_{\mathbf{k}}\left( l\right)$,
Eq. (\ref{Jk}), is still small.

\textit{Conductivity away from the MF transition point.} We can find
{rather easily} the temperature dependency of conductivity {in the
region}
\begin{equation}
l_{max}^{-1}\ll 1-J(0)/J_{c}\ll 1.  \label{nearcritrange}
\end{equation}%
{Eq.~(\ref{nearcritrange}) means that} the system is quite close to
the MF transition and the effective tunnelling amplitude
$J_{\mathbf{k}}\left( l\right) $ is strongly renormalized, although
being small. Deeper in the insulating regime, i.e. at smaller $J(0)$
than {that} in Eq.~(\ref{nearcritrange}), the renormalization is
negligible and so is the effect of {the cotunnelling} on the VRH.

Defining the characteristic radius $\zeta $ of the non-renormalized
intersite coupling as
\begin{equation*}
J_{\mathbf{k}}(0)\approx J(0)(1-\zeta^{2}\mathbf{k}^{2}),\quad \mathbf{k}%
\rightarrow 0
\end{equation*}%
we find from Eq.~(\ref{Jk}) the dependence of {the} tunnelling on
distance $|\mathbf{r}-\mathbf{r}^{\prime }|\gg \zeta $:
\begin{eqnarray}\label{Jofr}
    J_{\br\br^\prime}(l)=J(0)\frac{\xi^2}{\zeta^2}\int\frac{d\bk}{(2\pi)^d}
    \frac{e^{i\bk(\br-\br^\prime)}}{1+\xi^2k^2}
    \nonumber\\
    \sim
    J(0)\:\xi^{2-d}\zeta^{-2}|\br-\br^\prime|^{2-d}\exp({-|\br-\br^\prime|/\xi}),
\end{eqnarray}
where we defined the characteristic radius
\begin{eqnarray}
    \xi=\zeta\left[(\lambda J(0)l)^{-1}-1\right]^{-1/2}
    \label{xi}
\end{eqnarray}
of the renormalized coupling function.

At sufficiently low temperatures the RG flow should be stopped if
the characteristic number $Z(l)=E\nu\xi^d$ of states on length $\xi$
approaches unity. Eq.~(\ref{xi}) and $l=\ln (E_{0}/E)$ define
self-consistently the energy $E=(\nu \xi ^{d})^{-1}$ that terminates
the flow.

Thus, applying the RG procedure we arrive at a sparse array of sites
with the coordination number $Z$ of order unity, the characteristic
intersite spacing $\xi$, Eq.~(\ref{xi}), characteristic on-site
energies $E_{min}=1/(\nu \xi ^{d})$, and the tunnelling amplitudes
given by Eq.~(\ref{Jofr}). Inelastic hopping in the latter model is
dominated by the direct hops between the sites, while the
interference only slightly modifies the behaviour of the particle
wavefunctions.

Owing to the weakness of the tunnelling, each particle excitation is
almost entirely localized on one site. By construction of the
site-decimating RG the particle amplitude distance $r$ away from
this site estimates
\begin{equation}
    \Psi(r)\sim J_{{\bf 0}\br}[\ln(E_0/E_r)]/E_r,\,
    E_r=\max[(\nu r^d)^{-1},E_{min}]
    \label{Psi}
\end{equation}
in the first order in the small tunnelling. At large distances the
wavefunction decays exponentially $\propto\exp(-r/\xi)$, cf.
Eqs.~(\ref{Jofr}) and (\ref{Psi}). Applying the traditional Mott's
arguments\cite{Mott:book} we find the temperature dependence of
conductivity (Mott's law):
\begin{equation}
\sigma =\sigma _{0}\exp \left[ -\left( T_{M}^{\xi }/T\right) ^{1/(d+1)}%
\right] ,\: T\ll T_{M}^{\xi }\sim (\nu \xi ^{d})^{-1}, \label{sigma}
\end{equation}%
with a renormalized Mott's temperature $T_{M}^{\xi}$ matching the
value of the cutoff $E_{min}$ in the RG procedure.

At $T>T_{M}^{\xi }$ the RG should be stopped when the running energy
cutoff reaches temperature $T$, as the latter sets the
characteristic energy of conducting excitations. The power-lower
hopping probability on distances shorter than $\xi$ leads to the
temperature dependency of conductivity weaker than exponential. The
standard Mott's arguments yield\cite{VekilovIsaev} a power law
$\sigma \sim T^{\alpha }$, although those cannot be used to find
explicitly the power $\alpha $.

\textit{Effect of magnetic field.} In magnetic field each tunnelling
amplitude acquires a phase:
\begin{equation}
J_{\mathbf{r}_{1}\mathbf{r}_{2}}=|J_{\mathbf{r}_{1}\mathbf{r}_{2}}|\exp %
\left[ i{q}{c}^{-1}\int_{\mathbf{r}_{1}}^{\mathbf{r}_{2}}\mathbf{A}(\mathbf{r%
})d\mathbf{r}\right] .  \label{phase}
\end{equation}%
Here $q$ is the particle charge and the integration is carried out
along a certain path connecting sites $\mathbf{r}_{1}$ and
$\mathbf{r}_{2}$, which in general depends on the microscopic
details and for simplicity assumed below to be a straight line.
We neglect the modification of the non-renormalized values of the couplings $|J_{\mathbf{r}_{1}%
\mathbf{r}_{2}}|$ due to the \textquotedblleft
shrinkage of wavefunctions\textquotedblright \cite{Shklovskii:book}.
In principle, it can be straightforwardly included in the
magnetoconductance.

The RG procedure leaves the phase factors intact but modifies the
tunnelling amplitudes as
\begin{equation}
\partial _{l}|J_{\mathbf{r}\mathbf{r}^{\prime }}|=\lambda \int |J_{\mathbf{r}%
\mathbf{R}}||J_{\mathbf{R}\mathbf{r}^{\prime }}|\exp \left[ {iq}{c}%
^{-1}\oint_{\mathbf{r}\mathbf{R}\mathbf{r}^{\prime }}\mathbf{A}(\mathbf{r})d%
\mathbf{r}\right] d\mathbf{R}.  \label{RGA}
\end{equation}%
The contour integral in Eq.~(\ref{RGA}) gives the magnetic flux
through the triangular contour $\mathbf{r}\rightarrow
\mathbf{R}\rightarrow \mathbf{r}^{\prime }\rightarrow \mathbf{r}$.

Let us consider a two-dimensional system in a weak magnetic field $B$, so
that the renormalization is not suppressed and the length
\begin{equation}
L_{B}=\left[ c/(qB)\right] ^{1/2}
\end{equation}
exceeds the characteristic coupling radius $\xi$ in the absence of
magnetic field, Eq. (\ref{xi}). Then the RG should be stopped again
at the same high-energy cutoff $(\nu \xi ^{d})^{-1}$.

According to Eq.~(\ref{RGA}), the tunnelling on distance $r$ is
unaffected by magnetic field if the flux through the characteristic
``interference area''\cite{Perturbative} ${r}^{3/2}\xi ^{1/2}$ is
less than $c/q$, i.e. if ${r}^{3/2}\xi ^{1/2}<L_{B}^{2}$. Therefore,
the magnetic field has no effect on conductance provided the mean
inelastic hopping distance\cite{Mott:book} $\overline{r}\sim \xi
(T_{M}^{\xi }/T)^{1/3}$ does not exceed the scale $R\sim
L_{B}^{4/3}\xi ^{-1/3}$, which otherwise determines the
characteristic length of the inelastic hopping, as it serves as a
cutoff of the tunnelling $J_{\br\br^\prime}$ and the excitation
wavefunctions, Eq.~(\ref{Psi}).

\begin{figure}[t]
\includegraphics[width=.6\columnwidth]{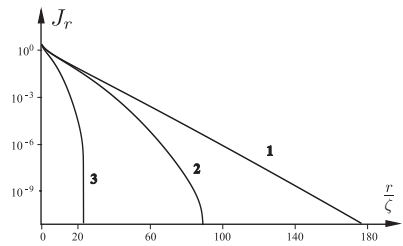}
\caption{The dependence of the renormalized tunnelling on distance,
obtained from the numerical solution of Eq.~(\ref{RGA}). Three
curves correspond to different values of magnetic field [in units
$c/(q\zeta^2)$] B= 1)0, 2)0.001, 3)0.01. $l=0.97l_{max}$.
\label{couplings}}
\end{figure}

At larger magnetic fields, $\xi >L_{B}$, the RG flow
should be stopped when the number of remaining states on length
$L_{B}$ approaches unity, and the renormalized coupling
is cut on the length $L_{B}$, Fig.~\ref{couplings}.
The conduction comes then from the
inelastic hopping on distance $R\sim L_{B}$ between sites of the
characteristic energy $1/(\nu L_{B}^{2})$.

\begin{figure}[t]
\includegraphics[width=.57\columnwidth]{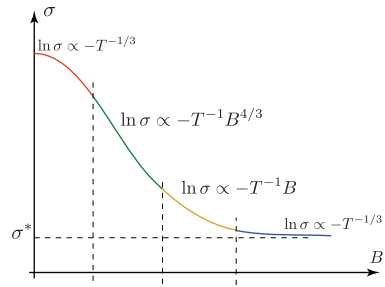}
\caption{(Colour online) The dependence of conductivity on magnetic
field in a two-dimensional system.
 \label{magnetoresistance} }
\end{figure}

Thus, we arrive at the conductivity
\begin{equation}
\sigma =\sigma _{0}\exp \left[ -\frac{1}{\nu T}\frac{\min (\xi
^{2/3},L_{B}^{2/3})}{L_{B}^{8/3}}\right]   \label{sigmaB}
\end{equation}%
that shows an activational behaviour with the activation gap $\propto
B$ in high magnetic fields and $\propto B^{4/3}$ in lower fields,
Fig.~\ref{magnetoresistance}. Upon decreasing the field so that
$T/T_{M}^{\xi }\gtrsim (\xi /L_{B})^{4}$ the
conductivity~(\ref{sigmaB}) crossovers to Eq.~(\ref{sigma}).

At very strong magnetic fields, $B\ggg c/(q\zeta^{2})$, the
interference between the tunnelling paths is destroyed, which
suppresses the renormalization and leads to the usual Mott's-law
conductivity, Eq.~(\ref{sigma}), with $\xi =\zeta$.

Recently a similar model of Cooper-pair hopping in an array of sites
with random energies and large coordination number $Z\gg1$ has been studied in
Refs.~\cite{Feigelman:rubbish, CayleySIT}
in order to describe the superconductor-insulator transition (SIT) in
disordered superconductive films. In Ref.~\cite{CayleySIT} the effective
 array of sites has been mapped
on a Bethe lattice and studied by means of an MF-type approach with
a site-dependent order parameter, using $Z\gg1$. We disagree with
such a mapping
 because Bethe lattice
contains no finite loops of sites and, thus, cannot account properly
for the interference phenomena considered here. Actually, the
mapping of Ref.~\cite{CayleySIT} cannot be justified even after
performing the renormalization, because the RG flow can be
stopped only when the characteristic coordination number $Z(l)$ reaches unity.

An interesting issue that deserves a separate investigation is the role of the intersite
interactions in the SIT in disordered superconductive films, as even
arbitrarily small long-range charging interactions are
known\cite{Firstorder} to renormalize the parameters of a clean
system and change the order of the phase transition.

\textit{Conclusion.} We considered low-temperature variable-range
hopping transport of bosons in an array of sites with random
energies. We constructed a renormalization group technique that
allows one to repeatedly eliminate the highest-energy on-site
states, renormalizing the intersite couplings between the
lower-energy ones. Using this procedure, we reduced the system to a
sparse array of sites and found the dependency of conductivity on
temperature and magnetic field, Eqs.~(\ref{sigma}) and
(\ref{sigmaB}). When increasing magnetic field the dependency
crossovers from the Mott's law to Arrhenius law with the activation
temperature proportional to the magnetic field.

\textit{Acknowledgements.} Our work has been supported by SFB
Transregio 12 and SFB 491. We have benefited from discussions with
L.B.~Ioffe, V.E.~Kravtsov, M.~M\"uller, B.I.~Shklovskii, and
B.~Spivak.


\end{document}